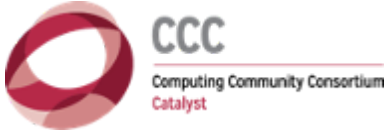

# Interdisciplinary Approaches to Understanding Artificial Intelligence's Impact on Society
*A Computing Community Consortium (CCC) Quadrennial Paper*

Suresh Venkatasubramanian (University of Utah), Nadya Bliss (Arizona State University), Helen Nissenbaum (Cornell University), and Melanie Moses (University of New Mexico)

### Overview

Long gone are the days when computing was the domain of technical experts. We live in a world where computing technology—especially artificial intelligence—permeates every aspect of our daily lives, playing a significant role in augmenting and even replacing human decision-making in a broad range of situations. AI-enabled technologies can adjust to your child's level of understanding by processing a pattern of mistakes; AI systems can leverage combinations of sensor inputs to choose and carry out braking actions in your car; web browsers with AI capabilities can reason from past observations of your searches to recommend a new cuisine in a new location.

Innovations in AI have focused primarily on the questions of "what" and "how"—algorithms for finding patterns in web searches, for instance—without adequate attention to the possible harms (such as privacy, bias, or manipulation) and without adequate consideration of the societal context in which these systems operate. In part, this is driven by incentives and forces in the tech industry, where a more product-driven focus tends to drown out broader reflective concerns about potential harms and misframings[1]. But this focus on what and how is largely a reflection of the engineering and mathematics-focused training in computer science, which emphasizes the building of tools and development of computational concepts. As a result of this tight technical focus, and the rapid, worldwide explosion in its use, AI has come with a storm of unanticipated socio-technical problems, ranging from algorithms that act in racially or gender-biased ways, get caught in feedback loops that perpetuate inequalities, or enable unprecedented behavioral monitoring surveillance that challenges the fundamental values of free, democratic societies.

Some of these problems are old ones in new forms; some—especially those wrought by the networked world of social media and the hidden ubiquitousness of AI technology in today's society—are entirely new. All of these problems follow from, and depend upon, the complex linkages between technology and society; they cannot be solved by narrow technological approaches. Rather, solutions will require a deeper recognition of how technology interacts with the people, as well as social institutions that it affects. This very *human and societal* understanding has long been the purview of the humanities and social sciences: disciplines that allow us to surface, formulate, probe, and examine questions about the underlying norms and values implied by the systems, methods, and tools of modern technology.

---

[1] Please see the companion CCC Quadrennial Papers entitled "A Research Ecosystem for Secure Computing" and "Next-Wave AI: Robust, Explainable, Adaptable, Ethical, and Accountable."

It is this critical perspective that has been absent from traditional computer-science training, research, and practice. And it is precisely this fundamental perspective that is needed in order to understand the consequences (good and bad) of technology for individuals and societies and how digital technologies are shaping our world. *Given that AI is no longer solely the domain of technologists but rather of society as a whole, we need tighter coupling of computer science and those disciplines that study society and societal values.* Initial efforts towards increased interaction between National Science Foundation's Computer and Information Science and Engineering (CISE) and Social, Behavioral, and Economic Sciences (SBE) directorates are promising. The Social Science Research Council's Just Tech program is another initiative funding research that will "imagine and create more equitable and representative technological futures". However, given the speed of both technical innovation and adoption of new technologies, it is vital for the federal government to broaden and deepen its support of these efforts in sustained, systematic ways.

**The Challenge**

Within the broader computer science community, many researchers (and especially those who study fairness, accountability and transparency) have called for a re-examination of the goals of the field from the perspective of society. This kind of self-examination is an integral part of the critical inquiry that is common in the social sciences, where one encounters clear, systematic, and rigorous thinking about ethical and societal values, and how particular framings of these concepts might be unintentionally laden with specific values.

Computer scientists are familiar with clear, systematic and rigorous thinking, but usually in the context of scientific and mathematical formalisms. To ensure that technical definitions map properly onto values that are both technically *and societally* appropriate—fairness, privacy, security, human autonomy, etc.—computing practitioners will need to work with humanists and social scientists to seek conceptual clarity for those key values. Moreover, they will need to make sure that that framing informs technological solutions from start to finish, since attempts to retrofit appropriate ethical structures on a deployed technology are rarely successful.

An informed, critical perspective asks us to reflect on whether the problems we have chosen to solve, and the ways in which we have chosen to define and address them, reflect an explicit set of value choices.  It has long been noted that technology is value-laden; it thus naturally follows that we should ask *which* values are being put forward when we build and use AI systems. A critical approach that brings together computing researchers with experts from the humanities and social sciences can help frame the AI enterprise in a manner that respects the values we seek to embody.

Many of the ethical problems we see with computing technology today can be traced to insufficient and inadequately informed attention to the configuration of values they embody. We call for an  approach that explicitly incorporates an understanding of and respect for the values in the milieu where the system will be deployed, and an awareness that these systems are often shared and re-deployed well beyond their original intentions. If we can give considerations of societal (ethical and political) values similar prominence as drivers such as performance, accuracy, and scalability, we may advance generations of technologies that are less likely to cause harm, and may even promote positive goals of social good—e.g., social justice, human autonomy, freedom from exploitation, and privacy.

This critical shift in the computer science perspectives and approaches cannot operate in isolation, nor can it proceed in separate disciplinary silos. Rather, computer science must look to fostering robust, ongoing communication and collaboration with relevant areas in the humanities and social sciences. This must not be merely a one-way street from these disciplines to computer science, or vice versa. The computing field has much to offer in the way we articulate problems, formalisms, limits on what is possible, and an imagination for what we can achieve. The humanities and social sciences can, as mentioned above, help frame and examine questions about norms and values. A true engagement between these different perspectives will require deep multi-disciplinary partnerships where researchers from different domains find common language, appreciate the different viewpoints that characterize each discipline, and craft solutions that respect values while also being technologically sound.

Creating such a culture will not be easy; the speed of innovation, incentives, drivers and timelines vary greatly between academia, industry, and the broader space of law and policy. Engagement with decision makers in government and policy circles will be essential to understanding the full landscape of the situation and the opportunities for solutions. And effective communication with this range of constituencies will be a major challenge.

### A Call to Action

To address the challenge of assessing, developing, and deploying digital technologies, including the potent products of AI, we need to create and sustain genuine multidisciplinary research partnerships. Experience suggests this is not easy: institutional and funding mechanisms are not organically constructed to promote and support these partnerships. Thus, not only will we need new funding initiatives but also new ways of *conceiving* such partnerships. A key piece of this new direction will be the design of evaluation and assessment approaches that will be effective in multidisciplinary settings.

By the time individuals have completed disciplinary training, they may already be trapped in structures that do not enable outreach or may even have narrowed perspectives. Thus, we see an urgent need for intervention at earlier stages, through curriculum development and training programs to provide coordinated exposure to perspectives relevant to partner disciplines, as well as community building to bridge the gaps. These are not changes that can be effected overnight, or on a project-by-project basis. Rather, they will require sustained, focused investment and alignment of organizational incentive structures to achieve, as outlined below.

*Mechanisms for transdisciplinary work.* Long-term, large-scale funding opportunities are common in computer science, engineering, and the natural sciences, encouraging new capabilities, methods, and systems. Humanist and social science initiatives tend to be smaller scale and focus on analysis, inquiry, and critique. An effective, sustainable research agenda for ethically and critically based computer-science research will require blending these communities and their approaches. One mechanism for doing so would be a significant (institute level) investment in large-scale research centers co-led by researchers from computing, the humanities, and the social sciences, working under a mandate to incorporate the ethical and critical perspective. These centers would draw in the associated science and engineering efforts in support of this work, creating significant reorientation of perspective and approach that would make a difference. Smaller-scale funding programs will also be important—again, with requirements for mixed teams, both at the leadership level and in the trenches, and diversity in the products of the work. These programs must be sustained, and time horizons for the awards must extend beyond the usual two or three-year term; changing a culture requires time.

*Evaluation and assessment.* Alongside these research initiatives, given the fundamentally different nature of evaluation in social science and humanities, it will be necessary to develop measures of effectiveness and progress for work that sits at their interface with computing. Those measures should be complementary to the traditional quantitative metrics that are typical of computer-science research (computational performance, precision, algorithm performance, etc). We must also consider how to develop metrics that can encourage researchers to work across disciplines. Such metrics must negotiate different publication cultures, venues for publishing, and methods for evaluating outreach and impact, both within the academy and in society at large. While clearly relevant to computing, social science, and humanities interdisciplinary work, this effort is likely to produce outcomes relevant for a broad range of interdisciplinary collaborations, where evaluation has proved to be a significant challenge for the research community.

*Education and training.* To develop a new generation of computer scientists who take a critical approach to socio-technical problems, it will be vital to create effective, sustained transdisciplinary education initiatives. These initiatives could be augmentations of existing training grant programs, but could also include interdisciplinary summer schools for Ph.D. students, cross-training fellowships for postdocs, and certificate programs for college students. We have seen great success with manageable efforts such as Summer Schools for Ph.D. students supporting participants from technical and non-technical fields, and in the case of technical students, to allow supported time and alleviate the need to seek industry-based internships. These models need to be sustainable over many years, if not decades. It will be important to ensure that these initiatives connect to community colleges, not just traditional research universities, and even to K-12 programs.

*Community building.* Finally, and consistent with development of a new set of metrics, we will need to prioritize support for initiatives that build community between computer scientists and their counterparts in complementary and related disciplines and amplify successes of collaborations in research, education, and networks.

**Summary**

Consistent integration of ethical and critical inquiry and other related methodologies ought to be an integral part of the disciplinary process in computer science, augmenting the traditional "what and if something can be done?" with "*should* it be done…and what potential vulnerabilities does it create?" Value-driven approaches make this possible, allowing technology to reflect the full racial, socioeconomic, gendered and political diversity of the society with which it will interact. We need to move away from the flawed idea that ethical thinking imposes limits on technical innovation. On the contrary, we view the integration of societal values into the creation of sophisticated digital systems as a way to spur innovation and deployment of these systems. A deeper engagement with the social sciences and humanities has enormous potential for productive reframing of basic questions in computation that will allow us to engender trust, do good, and do better.

*This white paper is part of a series of papers compiled every four years by the CCC Council and members of the computing research community to inform policymakers, community members and the public on important research opportunities in areas of national priority. The topics chosen represent areas of pressing national need spanning various subdisciplines of the computing research field. The white papers attempt to portray a comprehensive picture of the computing research field detailing potential research directions, challenges and recommendations.*

*This material is based upon work supported by the National Science Foundation under Grant No. 1734706. Any opinions, findings, and conclusions or recommendations expressed in this material are those of the authors and do not necessarily reflect the views of the National Science Foundation.*

*For citation use: Venkatasubramanian S., Bliss N., Nissenbaum H., & Moses M. (2020) Interdisciplinary Approaches to Understanding Artificial Intelligence's Impact on Society.*
*https://cra.org/ccc/resources/ccc-led-whitepapers/#2020-quadrennial-papers*